\documentclass[fleqn,11pt,a4paper]{wlscirep}

\usepackage{subfigure}
\usepackage{xcolor}

\title{Towards Printable Natural Dielectric Cloaks via Inverse Scattering Techniques}

\author[1,*]{Loreto Di Donato}
\author[2,$\dagger$]{Tommaso Isernia}
\author[3]{Giuseppe Labate}
\author[3,$\ddagger$]{Ladislau Matekovits}

\affil[1]{Department of Electrical, Electronics and Computer Engineering (DIEEI), University of Catania, Viale A. Doria 6, 95126, Catania, Italy.}

\affil[2]{Department of Information Engineering, Infrastructures and Sustainable Energy (DIIES), University ``Mediterranea'' di Reggio Calabria, Via Graziella, Loc. Feo di Vito, 89100, Reggio Calabria, Italy.}

\affil[3]{Department of Electronic and Telecommunications (DET), Politecnico di Torino, C.so Duca degli Abruzzi 24, 10129,
Torino, Italy}

\affil[$\dagger$]{also with Consorzio Nazionale Interuniversitario per le Telecomunicazioni, 43124, Parma, Italy.}

\affil[$\ddagger$]{also with the Macquarie University, 2109, Sydney NSW, Australia}

\affil[*]{loreto.didonato@dieei.unict.it}

\begin{abstract}
The  synthesis of non-magnetic 2D dielectric cloaks as proper solutions of an inverse scattering problem is addressed in this paper. Adopting the relevant integral formulation governing  the scattering phenomena, analytic and numerical approaches are exploited to provide new insights on how frequency and direction of arrival of the incoming wave may influence the cloaking mechanism in terms of permittivity distribution within the cover region.
In quasi-static (subwavelength) regime a solution is analytically derived in terms of \textit{homogeneous} artificial dielectric cover with $\varepsilon<\varepsilon_0$ which is found to be a necessary and sufficient condition for achieving \textit{omnidirectional} cloaking. On the other hand, beyond quasi-static regime, the cloaking problem is addressed as an optimization task looking for only natural dielectric coatings with $\varepsilon>\varepsilon_0$ able to hide the object for a given number of directions of the incident field. Simulated results confirm the validity of both analytic and numerical methodologies and allow to estimate effective bandwidths both in terms of frequency range and direction of arrival of the impinging field. 
\end{abstract}

\begin{document}
%
\maketitle
%
%
\thispagestyle{empty}
\noindent 
%
%
\section*{Introduction}
The possibility to retrieve the shape and constitutive parameters of a 
medium from its scattered field is known as  \textit{detection} and/or \textit{imaging} problem and it is related to proper solutions of an electromagnetic inverse scattering problem (ISP) \cite{Colton}. In this respect, the opposite task is avoiding such detection by hiding scatterers from external observers: this is known in the literature as \textit{cloaking} and can occur, for example, exploiting plasmonic and metamaterial coatings
\cite{PC,TO}.

Negative, or less than unity, $\varepsilon$ (ENG or ENZ) and $\mu$ (MNG or MNZ) materials \cite{Brown} serve to induce cancellation effects on scattered fields as in Plasmonic Cloaking  (PC) \cite{PC} or to reroute the impinging waves as in Transformation Optics (TO) \cite{TO}. The common idea is acting on the internal field within or through the hidden region without perturbing the surrounding space. 
However, PC shows two main limitations: it requires plasmonic materials and formulas are valid only under quasi-static regime. For these reasons, the method is not well suited for objects that are large or comparable with respect to the impinging wavelength, especially in the radiofrequency and microwave bands where ENG and ENZ materials are not seldom available. On the other hand, by construction, TO requires inhomogeneous and strongly anisotropic dielectric and magnetic tensors to practically manufacture the cloak. Indeed, the quest for an exact, object-independent, invisibility coating is paid by enormous complications (if not impossibility) of the desired cloak \cite{Wolf}.

Moreover, the possibility to achieve cloaking via PC or TO in actual applications is severely limited by the difficulty of considering the influence of the impinging wave (coming from a scanned array for example) on devices with a low scattering response quite sensitive to the working frequency and direction of arrival (DoA) of the incident field. 

In this paper, we address the synthesis of dielectric covers able to achieve null or very low values of the scattered fields (cloaking effect) wherein only natural dielectric materials (i.e., without negative or near-zero values) without magnetic properties ($\mu=\mu_0$) are considered.
In the relevant literature, non-magnetic cloaks have been investigated for near-perfect invisibility \cite{Galdi} via TO. The use  of only natural dielectric materials have been addressed by means of topology optimization approaches \cite{Sigmund1,Sigmund2,Smith} at optical frequencies in the case of metallic objects, requiring extremely variable covers to induce cloaking exclusively for a narrow angular range of DoAs. Other all-ordinary dielectric cloaks have been proposed via global optimization techniques for radially cylindrical and spherical cloaks of metallic targets \cite{Kishvar,Wang1,Wang2,Yu,Ludutenko}. These approaches, although very attractive in terms of simplicity of the geometrical architecture of the covers, require very large refractive indexes \cite{Wang1,Wang2}, rarely available in nature, or even near-zero dielectric constants \cite{Yu}, moreover they achieve only nearly optimal performance in terms of residual scattering radiation \cite{Ludutenko}. Another optimization approach based on the \emph{phase field method} has been recently proposed \cite{Heo2016} for cloaking metallic cylinder, by considering six angular directions of the incident wave. Finally, graded refractive index structures have been proposed for surface cloaking \cite{Valentine2009,LaSpada2016}.

Differently from the above contributions mostly concerned with cloaking metallic objects and surface, in this paper, we investigate through recent analytic \cite{Laby1} and numerical methods \cite{Loreto}, the conditions to pursue non-magnetic volumetric dielectric cloaking with natural permittivity materials ($\varepsilon > \varepsilon_0$) to cloak dielectric objects. In this respect, we adopt a design procedure based on the solution of the inverse scattering problem \cite{Colton} exploiting a gradient based local optimization approach which allows to easily handle many of the design constrains concerned with the cloaking problem. 
A physical insight into \emph{artificial-natural} and \emph{natural-natural} dielectric cloaking systems is also given. The presented methodology is mainly focused on investigations against volumetric dielectric scatterers, with a possible extension towards composite structures such as metal-dielectric metasurfaces \cite{Mat}.

\section*{Methods: Analytic and Numerical Cloak Synthesis Procedures}\label{sec:2}
\begin{figure}
\centering
\includegraphics[width=0.50\textwidth]{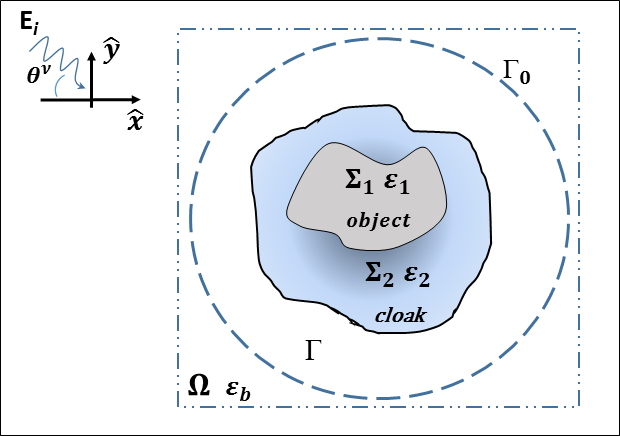}
\caption{\small Geometry of the problem or the synthesis of dielectric cloaking: $\Omega$ is the computational domain which includes arbitrarily shaped scattering system $\Sigma$ (cloaking region) and $\Gamma=\Omega/\Sigma$ (observation region). The $\Sigma$ region, divided into $\Sigma_1$ (bare object) and $\Sigma_2$ (cloak cover), is illuminated by an incident field $E_i$ impinging from an angular directions $\theta_\nu$. $\Gamma_0$ is a subset of $\Gamma$ where the scattered field is enforced to be zero in the design procedure based on inverse scattering technique.} \label{fig:scenario}
\end{figure}
We consider a 2D domain $\Omega$ embedding one (or more) penetrable non-magnetic homogeneous object(s) of arbitrary shape with support $\Sigma_1$, see Fig.\ref{fig:scenario}. A cloak region $\Sigma_2$ such that $\Sigma=\Sigma_1 \cup \Sigma_2$ is considered, and, for the sake of simplicity, the intersection between $\Sigma_1$ and $\Sigma_2$ form a null set. However, the analysis developed in the following is valid even if this hypothesis is removed.

Let us assume the $\Omega$ domain in the $xy$ plane, and one (or more) plane wave(s) 
\begin{align}
\underline{E}_i( \underline{r},\theta_\nu, \omega)=e^{- j \underline{k} (\omega,\theta_\nu) \cdot \underline{r}} \;  \hat{z}
\end{align}
with unitary amplitude and electric field polarized along the $\hat{z}$ axis (TM polarization) impinging towards the center of $\Omega$, wherein $\underline{r}=(x,y)$ denotes the vector position in $\Omega$, $\omega=2\pi f$ the angular frequency and $\theta_{\nu}$ the DoA. Throughout the paper, the time harmonic factor $e^{j\omega t}$ is assumed and dropped.

For the sake of clarity, the dielectric properties of the overall region are grouped in:
\begin{align} \label{eq:eps_con}
& \varepsilon(\underline{r})=
\left\{
\begin{array}{ll}
    \varepsilon_{1}(\underline{r}), \; \underline{r}\in \Sigma_1 & (\text{object region}) \\
    \varepsilon_{2}(\underline{r}), \; \underline{r}\in \Sigma_2 & (\text{cloak region}) \\
    \varepsilon_b \; \hspace{4mm}, \; \underline{r}\in \Gamma & (\text{observation region})
\end{array}%
\right.
\end{align}
with $\Gamma\equiv\Omega/\Sigma$ the observation region where no ``visible'' scattering effects by the cloaking system would be desired and where the constitutive parameters are homogeneous (i.e., $\varepsilon_b$ does not depend on $\underline{r}$ in $\Gamma$).

From now on, arrow signatures on fields are suppressed, tacitly assumed all vectors being directed along $\hat{z}$. 
The electromagnetic scattering from such a cylindrical (i.e., longitudinal invariant) structure is due to the equivalent volumetric sources with support $\Sigma$, defined as $J_{eq}(\underline{r},\theta_\nu, \omega)= j \omega \varepsilon_b  J(\underline{r},\theta_\nu, \omega)$ where
\begin{equation}\label{eq:jeq}
 J(\underline{r},\theta_\nu, \omega)=
\chi(\underline{r})\begin{bmatrix}E_i(\underline{r},\theta_\nu, \omega) + E_s(\underline{r},\theta_\nu, \omega) \end{bmatrix}
\end{equation}
In Eq. \eqref{eq:jeq}, $J(\cdot)$  is the so called \textit{contrast source} given by 
the product between total internal fields $E_t(\underline{r},\cdot)\equiv E_i(\underline{r},\cdot)+E_s(\underline{r},\cdot)$ and $\chi(\underline{r})$ is the contrast function defined as:
\begin{equation}\label{eq:chi}
\chi(\underline{r})=\frac{\varepsilon(\underline{r})-\varepsilon_b}{\varepsilon_b}.
\end{equation}
The total field $E_t$, expressed in the whole region $\Omega$ as the sum of the incident (or primary) and scattered (or secondary) field, can be conveniently expressed via integral formulation as \cite{Richmond}:
\begin{equation}\label{data}
E_t(\underline{r},\theta_\nu,\omega)=E_i(\underline{r},\theta_\nu, \omega) +k_b^2
\iint\limits_{\Omega} J(\underline{r}',\theta_\nu, \omega)
G(\underline{r},\underline{r}', \omega)d\underline{r}'
\end{equation}
where $k_b=\omega \sqrt{\mu_b\varepsilon_b}$ (recalling that everywhere $\mu_b=\mu_0$).
Moreover $G(\underline{r},\underline{r}',\omega)$ is the 2D Green's function of the homogeneous background which has exact analytic form in a homogeneous background, namely the Hankel function of zero order and second kind  \cite{Richmond}.
According to the partition in Eq.\eqref{eq:eps_con}, by definition, the contrast function is zero in the observation domain $\Gamma\equiv\Omega/\Sigma$ and it is mostly non-zero elsewhere. Let us also notice that, by definition, if vacuum is assumed as background medium (i.e., $\varepsilon_b=\varepsilon_0$), the contrast function is coincident with the electric susceptibility. 
Eq.\eqref{data} states that the physical cause of the scattering phenomenon is
the contrast source $J$ induced in $\Sigma$. It may be convenient  to express the scattered field through a more compact notation with respect to $\Sigma$ and $\Gamma$ domains, as:
\begin{align}
\label{eq:int}
& E_t(\underline{r},\theta_\nu, \omega)-E_i(\underline{r},\theta_\nu, \omega)= \mathcal{A}_{\Sigma}[J] \hspace{8.5mm}\mbox{with $\underline{r} \in \Sigma$} \\
& E_s(\underline{r},\theta_\nu, \omega)=\mathcal{A}_{\Gamma}[J]
\hspace{28.5mm} \mbox{with $\underline{r} \in \Gamma$} \label{eq:ext}
\end{align}
Eqs. \eqref{eq:int} and \eqref{eq:ext} can be identified as the \textit{object} and \textit{data} integral equations of the ISP, respectively. From a physical point of view, the operators   $\mathcal{A}_{\Sigma}$ : $L^2(\Sigma) \rightarrow   L^2(\Sigma)$  and $\mathcal{A}_{\Gamma}$ : $L^2(\Sigma) \rightarrow  L^2(\Gamma)$ map the contrast source into the corresponding scattered field in $\Sigma$ and in $\Gamma$, respectively.
Adopting this formulation, the cloaking effect can be pursued by enforcing 
\begin{align}
E_s(\underline{r}, \theta_\nu, \omega)=0 \hspace{5mm}  \forall   \mbox{ $\underline{r}\in \Gamma$}.
\label{eq:ideal_cloak}
\end{align}
In the following, solutions of Eqs. \eqref{eq:int}-\eqref{eq:ext} are pursued to achieve cloaking in and beyond the quasi-static regime, using both analytic and numerical approaches.
\subsection*{Cloaking in quasi-static regime}\label{sec:2.1}
When Eq. \eqref{eq:ideal_cloak} is assumed as desired specification of any cloaked system, the data equation \eqref{eq:ext} can be explicited as:
\begin{align}
\iint\limits_{\Sigma} J(\underline{r}',\theta_\nu,\omega)
G(\underline{r},\underline{r}',\omega)d\underline{r}'
=0 \hspace{5mm}   \mbox{with $\underline{r}\in \Gamma$}.
\label{eq:solve_cloaking}
\end{align}
Considering the overall system enclosed in a circular cylinder of radius $b$, in quasi-static condition (i.e., $k_b b \rightarrow 0$), Eq. \eqref{eq:solve_cloaking}  can be solved in a straightforward and simple manner since the overall system is so extremely compact in terms of $\lambda$ that the Green’s function of the homogeneous background becomes constant over the entire domain $\Sigma$, i.e.,
\begin{align}
\lim_{k_b b \rightarrow 0} G(\underline{r},\underline{r}',\omega)= \lim_{k_b b \rightarrow 0}-\dfrac{j}{4} \mbox{H}_0^{(2)}(k_b|\underline{r}-\underline{r}'|)= C \hspace{7mm} 
\end{align}
since $k_b|\underline{r}-\underline{r}'|\approx 2\pi b/\lambda$ in the quasi static-limit.
Expliciting the contrast source according to Eq. \eqref{eq:jeq}, also the total field can be considered to be constant in the $\Sigma$ domain in the quasi-static approximation, i.e., the Rayleigh scattering regime, thus giving:
\begin{equation}\label{CCE}
\iint\limits_{\Sigma}\chi(\underline{r}')  d\underline{r}'=0.
\end{equation}
Splitting Eq.\eqref{CCE}, namely the \textit{Contrast Cloaking Equation} (CCE) \cite{Laby1} over  the object region, $\chi_1 \in \Sigma_1$ and the cloak region $\chi_2 \in \Sigma_2$, a necessary and sufficient condition to achieve cloaking comes out as a proper mix of positive/negative values of the local contrast function, i.e.,
\begin{equation}\label{CCE_algebraic}
\chi_1 \Sigma_1+\chi_2\Sigma_2=0.
\end{equation}
It is interesting to notice that the CCE generalizes in a very compact fashion PC for scatterers of arbitrary shape \cite{Miano} for a general background medium, even when $\varepsilon_b \neq\varepsilon_0$. Therefore, in the quasi-static limit, the designer can turn-off the effect of volumetric sources by locally compensating the positive-induced source associated to a
positive contrast (e.g., the object) with the negative-induced one associated to a negative contrast (e.g., the cloak) or viceversa, regardless shape of the cloak system and DoAs of the incident field. As a matter of fact, since the total field does not play any role in this derivation (being factored out from the integral), such a kind of cloaking is expected to behave as an omni-directional cloaking, i.e., its performances do not depend from the DoA of the incident field.

\subsection*{Cloaking beyond quasi-static regime}\label{sec:2.2}
When the overall cloaking dimension is not in subwavelength condition, the two main hypothesis for deriving the
CCE are no longer valid. Beyond quasi-static condition,  distributed effects take place with two consequences: ($i$)
the Green's function is no more constant and ($ii$) the total field changes
from point to point, taking into account all the non-local contributions of the scatterers in the whole domain $\Sigma$ \cite{Richmond}.

\begin{figure}
\centering
\includegraphics[width=0.65\textwidth]{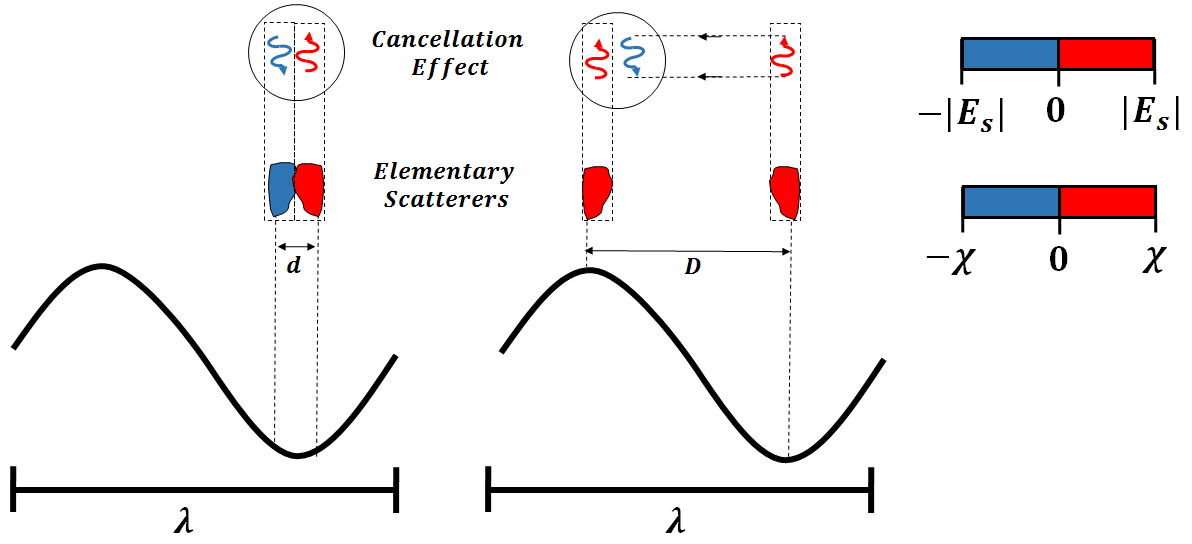}
\caption{\small (Left) Scattering cancellation with positive (red) and negative (blue) contrast $\chi$ at subwavelength scale $d/\lambda \approx 0$ as a necessary and sufficient condition for lumped elements. (Right) Beyond quasi-static regime, complete positive $\chi$ values (e.g., at a distance  $D/\lambda \approx 0.5$) can support the same effects on the local scattered field ($\pm |E_s|$). When properly designed, all the local distributed natural materials can achieve zero scattered field at any point outside the domain $\Sigma$.} \label{fig:ISPcloaks}
\end{figure}

However, thanks to these considerations, as also depicted in Fig. \ref{fig:ISPcloaks}, we figure out an important finding: the cloaking mechanism can be achieved with \textit{all-positive} values of the contrast provided that a proper spatial organization of the contrast layout is pursued. As a result, the cancellation effects occurring between positive-negative values of the contrast, that are very close at subwavelength scale, can be synthesized also for all-positive contrast values which are not so close in terms of wavelength (e.g., crest and through of the working wavelength). On the other hand, due to the need of specific spatial distribution of the contrast function, the cloaking effect is expected not to be broadband. 
As a matter of fact, when the ratio $D/\lambda$ increases, $D$ being the diameter of the minimum circle enclosing $\Sigma$, the architecture scheme of the coating plays a crucial role with the possibility to get rid of
metamaterials in $\Sigma_2$  through a proper arrangement of all-positive dielectric values $\varepsilon_2(\underline{r})  \geq \varepsilon_0$. On the other hand, as stressed above,  according to PC \cite{PC},  just one homogeneous layer of metamaterial is sufficient to achieve cancellation effects in quasi-static regime.

Under the above reasonings, the synthesis of cloaking profiles can be conveniently addressed as the solution of an ISP without any approximation on the mathematical model in Eqs. \eqref{eq:int}-\eqref{eq:ext}. This can be performed through the minimization of a proper cost function, under the constraint $\chi_2(\underline{r})>0$. Obviously, solutions accounting for both positive and negative values of the contrast function can be anyway pursued, but their investigation is beyond the scope of the present paper.
Once fixed $\chi_1$ in $\Sigma_1$, the adopted formulation of the cost function depends on both contrast $\chi_2(\underline{r})$ $\forall$ $\underline{r} \in \Sigma_2$ and contrast source $J(\underline{r})$ $\forall$ $\underline{r} \in \Sigma$. In particular, the cost function is obtained
as joint minimization of the weighted object  and data equation \eqref{eq:int}-\eqref{eq:ext}. To this aim, it is convenient to modify the  object equation according to the contrast source formulation\cite{Abubakar}, just  multiplying both members of Eq. \eqref{eq:int} by $\chi$. Moreover, the part of functional \eqref{cost_function} relevant to the data equation has to be properly modified to take into account zero scattered field enforced in the optimization task. A possible way to address such a problem is given by the minimization of the following weighted cost function:
\begin{align}
\Phi(\chi,J)=\sum_{\nu=1}^{N} \biggl\{ {\dfrac{||J^\nu-\chi E_i^\nu-\chi  \mathcal{A}_{\Sigma}[J^\nu]||^2}{||E_i^\nu||^2}}+
{\dfrac{||\mathcal{A}_{\Gamma_0} [J^\nu]||^2}{||E_i^\nu||^2}} \biggr\} \;\;\;  \nu=1,2.., N_\nu,
\label{cost_function}
\end{align}
\begin{figure}[hb!]
\centering
\includegraphics[width=0.75\textwidth]{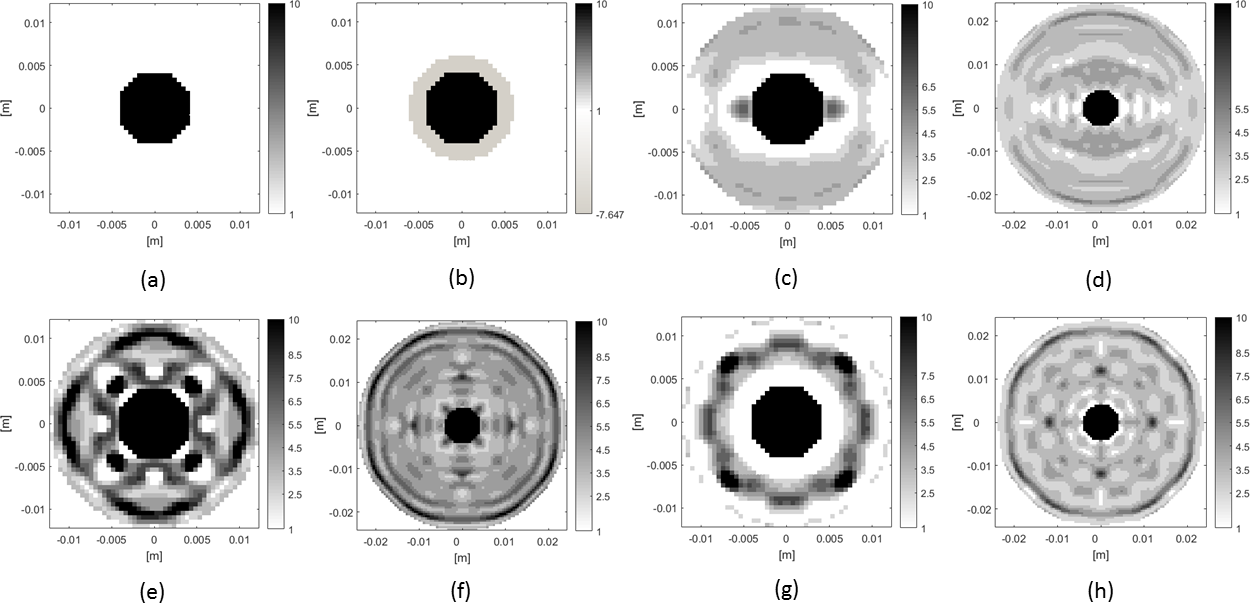}
\caption{\small Permittivity distribution for synthesized covers to cloak allumina disk: 
(a) bare object;
(b) homogeneous plasmonic cover at subwavelenght regime (f.i., $4$ GHz); 
(c)-(h) ordinary all dielectric cloaks operating at $25$ GHz with different features specified in Table \ref{tab:cloak_ISP}.}
\label{fig:cloak_cover}
\end{figure}
\begin{table}[hb!]
\begin{tabular}{cccccc}
\toprule
 CASE & $f_c$ & $b$ & $\varepsilon_2$ (or starting guess) & DoAs & $\Phi$  \\ 
\toprule 
(b) Plasm-cloak  & $4$ GHz & $0.16\lambda$ & $\varepsilon_2=-7.64 \varepsilon_0$ & Omnidirectional & Analytical Synthesis \\
(c) Diel-cloak  & $25$ GHz & $1\lambda$ & $\varepsilon_2=+3.50 \varepsilon_0$ & $\theta_\nu=k\pi$ with $k=0,1$ & $\Phi= 9.98\cdot 10^{-7}$ \\
(d) Diel-cloak  & $25$ GHz & $2\lambda$ & $\varepsilon_2=+3.50 \varepsilon_0$ & $\theta_\nu=k\pi$ with $k=0,1$ & $\Phi= 9.98\cdot 10^{-7}$ \\
(e) Diel-cloak  & $25$ GHz & $1\lambda$ & $\varepsilon_2=+5.50 \varepsilon_0$ & $\theta_\nu=k\pi/2$ with $k=0,..,3$ & $\Phi= 2.15\cdot 10^{-5}$ \\
(f) Diel-cloak  & $25$ GHz & $2\lambda$ & $\varepsilon_2=+5.50 \varepsilon_0$ & $\theta_\nu=k\pi/2$ with $k=0,..,3$ & $\Phi= 7.33\cdot 10^{-5}$ \\
(g) Diel-cloak  & $25$ GHz & $1\lambda$ & $\varepsilon_2=+4.50 \varepsilon_0$ & $\theta_\nu=k\pi/4$ with $k=0,..,7$ & $\Phi= 8.84\cdot 10^{-3}$ \\
(h) Diel-cloak  & $25$ GHz & $2\lambda$ & $\varepsilon_2=+4.50 \varepsilon_0$ & $\theta_\nu=k\pi/4$ with $k=0,..,7$ & $\Phi= 3.46\cdot 10^{-5}$ \\
\bottomrule
\end{tabular}
\centering
\caption{\small Plasmonic Cloak operating at $4$ GHz  (b) and ordinary all dielectric cloaks (c)-(h) operating at $25$ GHz with detailed parameters of the ISP design procedure.}
\label{tab:cloak_ISP}
\end{table}
\begin{figure}[ht!]
\centering
\includegraphics[width=0.75\textwidth]{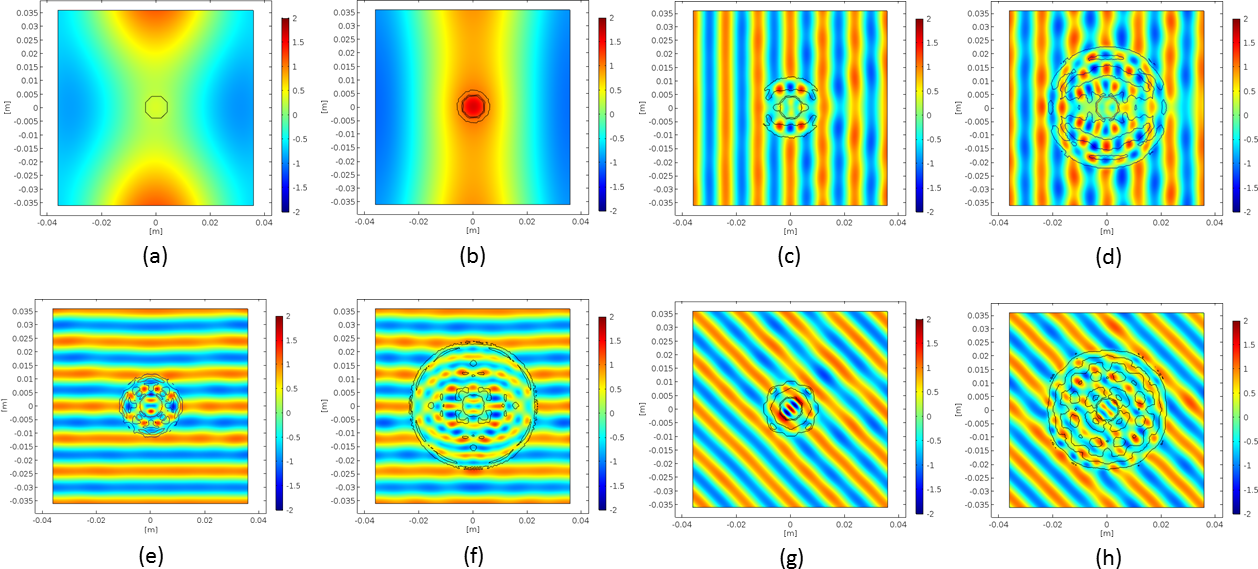}
\caption{\small  Real part of the total electric field for the bare and cloaked allumina disk following the order reported in Fig. \ref{fig:cloak_cover}: (a)-(b) $4$ GHz and (c)-(h) $25$ GHz.}
\label{fig:cloak_field}
\end{figure}
where the apex $^\nu$ stands for the ${\nu-th}$ impinging DoA and $|| \cdot ||$ denotes the usual $L^2$-norm. In particular, in Eq. \eqref{cost_function}, the first addendum enforces, for a given set of plane waves, the solution (in the least square sense) of
the scattering equation in $\Sigma$, while, interestingly, the second term stands for the minimization of the radar cross section (RCS), or the echo width \cite{Richmond} in a subset $\Gamma_0 \in \Gamma$, see Fig. \ref{fig:scenario}. In this respect, it is worth noting that the finite bandwidth of the scattered fields \cite{Buc_Is} allows to not enforce zero scattered field in the whole observation region $\Gamma$, since it is sufficient to enforce such a value only in a finite number of points $M$ over a surface $\Gamma_0$. For circular cylindrical geometry (a circumference of radius $R$ for the particular case at hand), according to \cite{Buc_Is}, the minimum non redundant number of sampling points $M=2k_b R$ can be considered as angularly equispaced along the circumference $\Gamma_0$ enclosing the cloaking system.

As a last comment, it is worth noting that the minimization problem in Eq. \eqref{cost_function} entails a non-quadratic form \cite{Isernia1997,Loreto}, so that the minimization procedure may get stuck into ``local minima'' \cite{Isernia2001}. The possibility to incur in local minima is strictly related to the functional shape (depending on the data and constraints of the problem) as well as on the \textit{initial guess} adopted to start the minimization procedure. In this respect, while the first set of parameters are fixed in advance by the designer, the initial guess is a ``degree of freedom'' of the synthesis problem, which can be conveniently exploited to obtain several equivalent (in terms of cloaking effect) solutions provided that a satisfactory weighted residual error is reached in the minimization of \eqref{cost_function}. 
\section*{Numerical Results and Analysis}
\label{sec:3}
To keep things simple, we have considered the cloaking of a circular scatterer in free-space background (i.e., $\varepsilon_b=\varepsilon_0$), made up of  lossless allumina ($\varepsilon_{s}=10\varepsilon_0$) with radius $a=0.42$ cm, see Fig. \ref{fig:cloak_cover}(a). However, it is worth noticing that such an approach can be easily adopted for any arbitrary (more sophisticated) shape of the scattering system (object and cover). 

The cloaking effect will be demonstrated, both in and beyond the quasi static-limit, adopting circular coatings with different radii. 

To achieve cloaking in the subwavelength condition, we set the cover's radius to $b=0.6$ cm (slightly larger than that of the bare object). In particular, the coating diameter $2b$ corresponds to $0.16\lambda$ at $4$ GHz (quasi-static regime).

In quasi-static condition the CCE is exploited and, following Eq. \eqref{CCE_algebraic}, the contrast in $\Sigma_2$ is computed as $\chi_2=-\chi_1 \Sigma_1/\Sigma_2$, where $\Sigma_1= \pi a^2$ and $\Sigma_2=\pi b^2-\Sigma_1$. The resulting value is $\chi_2=-8.64$ (i.e., $\varepsilon_2=-7.64\varepsilon_0$) and the overall cloaking system is reported in Fig. \ref{fig:cloak_cover}(b). 

Beyond subwavelength condition, the architecture for the natural dielectric cloak is obtained solving the optimization problem \eqref{cost_function} via conjugate gradient fast Fourier transform method (CG-FFT) \cite{Isernia1997} adopting a standard pixel representation for the discretization of the state and data equations, for both the scatterer and the cover $\varepsilon_2(\underline{r})$ in $\Sigma_2$. Two different dimensions for the cloak have been considered in order to take into account the performance of the cloak with respect to its electrical dimension (compared to the operational wavelength). The discretization of the analysis domain has been fixed according to rules of the integral equation method (MoM) \cite{Richmond}. Specifically, the dimension of the pixel has been set to $0.5$mm, which is slightly larger than $\lambda/10$ of the minimum wavelength (as referred to the allumina). This choice has been established as trade-off between the accuracy required by the numerical procedure and the need to deal with reasonable resolution in realizing the cloak trough solid printing techniques. 

In order to take into account the dependence of the cloak from the impinging directions of the incoming wave, we have considered in the optimization problem \eqref{cost_function}, different  number of DoAs, i.e. $N=2$, $N=4$ and $N=8$ plane waves angularly evenly spaced on a arc of $360^{\circ}$. On the other hand, since the cloaking effects is required all around the cloak we have enforced the scattered field to be zero in $M=24$ or $M=36$ equispaced observation points located on a circumference placed in the close proximity of the cloaking system. The circumference's radius has been set to $R \simeq 1.45 b$, $b$ being the different coating radius, with $b=1.2$ cm and $b=2.4$ cm ($1\lambda$ and $2 \lambda$ at 25 GHz, respectively) adopted in the synthesis procedure. These different values have been considered in order to take into account the different number of degrees of freedom of the scattered field pertaining to scatterers of different extent \cite{Isernia1997}.

According to the above reasons, we have considered a discretization grid of $48\times48$ pixels for the cloak with radius $b=1.2$ cm ($1\lambda$) and $96\times96$ pixels for the cloak with radius $b=2.4$ cm ($2\lambda$). Moreover, in the minimization procedure, constraints about natural permittivity (i.e., $\varepsilon \geq \varepsilon_0$) and lossless (i.e., $\sigma=0$) nature of the cover have been enforced at each step of the minimization procedure. In this respect, the maximum value of the permittivity to be used in the cover region has been set to that of the allumina (i.e., $10 \varepsilon_0$). It is worth noticing this constrain simplifies the fabrication of the cloak in terms of number of materials to be employed, as well as, not less important, allows to take under control, i.e. to not violate, the spatial discretization of the integral equations involved in the synthesis strategy. 

The synthesis procedures was stopped when the functional reached a value of $\Phi \le 10^{-6}$. This ensures that both data and state equation are solved with a satisfactory accuracy. On the other hand, when the threshold value was not reached, the procedure has been stopped when no significant changes arose in the functional value between two subsequent steps of the minimization procedure. 
 
The initial guess of the dielectric cover has been set as homogeneous in the whole $\Sigma_2$ with relative permittivity of $\varepsilon_2 =3.5\varepsilon_0$ or $\varepsilon_2 =5.5\varepsilon_0$, for $N=2, 8$ and $N=4$, respectively, in order to escape from local minima wherein the residual value of the functional \eqref {cost_function} is unsatisfactory for achieving cloaking effects. It is worth to note that these initial values have been chosen after some a posteriori checks.
 
The results of the synthesis strategy are shown in Fig.  \ref{fig:cloak_cover}(c)-(h), after a proper ``post-quantization'' procedure, which slightly adjusts the dielectric profile of the cover to permittivity's levels  with of $\varepsilon_r={1, 2.5,3.5,4.5,5.5,6.5,7.5,8.5,10}$. Such procedure entails a simplification of the cloak manufacturing in terms of piecewise constant dielectric profile taking into account the availability of solid materials in the considered frequency range, while not substantially modifying the cloaking performance, which results only slightly worsened. 
Fig. \ref{fig:cloak_field}(a) and \ref{fig:cloak_field}(b) show the real part of the total field for the uncloaked case and the plasmonic coating at $4$ GHz: in Fig. \ref{fig:cloak_field}(c)-(h),  the real part of the total field for all the natural dielectric cloaks at $25$ GHz is reported. The analysis was performed in COMSOL importing the synthesized profile and meshing it by means of triangular shaped element. 

 
At the lower frequency, the plasmonic cover is in cloaking operation with the flat phase fronts well recognizable behind the object, whereas, as expected, at the same frequency, no scattering cancellations occur for the natural dielectric cover (not shown).
At frequency of 25GHz, the dielectric cloaks are all able to achieve cloaking mechanism for the DoAs considered in the design procedure (for the sake of brevity the cloaking effect is shown only for one impinging directions). From these results it is possible to observe that the cloaking can been achieved both for the cloak with radius of $1\lambda$ and $2\lambda$, and that the smaller the cloak and the larger the number of DoAs, the more complicated the architecture of the covers to be synthesized is. Moreover, note that not all the discrete permittivity's values reported above are required by the synthesized covers. For example, for the cloaks of Fig.  \ref{fig:cloak_cover} (c) and (h) only four kind of dielectrics are required with a further simplification in the architecture of the cloak.

In order to quantify the cloaking effect as a function of the frequency, we have calculated the scattering cross section (SCS) of the synthesized cloak, which is the average value of the RCS calculated at $72$ equispaced points on a circle of radius $R=1.45 b$  for each given DoA (the same shown in Fig. \ref{fig:cloak_field}) \footnote{The numerical values of the SCS when calculated for different DoAs do not substantially change.}. 
\begin{align}
SCS(\theta_\nu,\omega)= 2\pi R \left[ \frac{1}{M} \sum_{m=1}^M  \left| \frac{E_{s}(\theta_\nu,\omega)}{E_{i}(\theta_\nu,\omega)} \right| ^{2}  \right] \
\;\;\;\;\; 
\label{C_gain}
\end{align}
The overall function SCS$(\theta_\nu,\omega)$ has been calculated for different frequency values at a fixed DoA and for different DoA values at a fixed (designed) frequency: this leads to quantify the bandwidth performance of the synthesized cloaks in terms of frequency and omnidirectionality issue.
\begin{figure}[ht!]
\centering 
\includegraphics[width=0.65\textwidth]{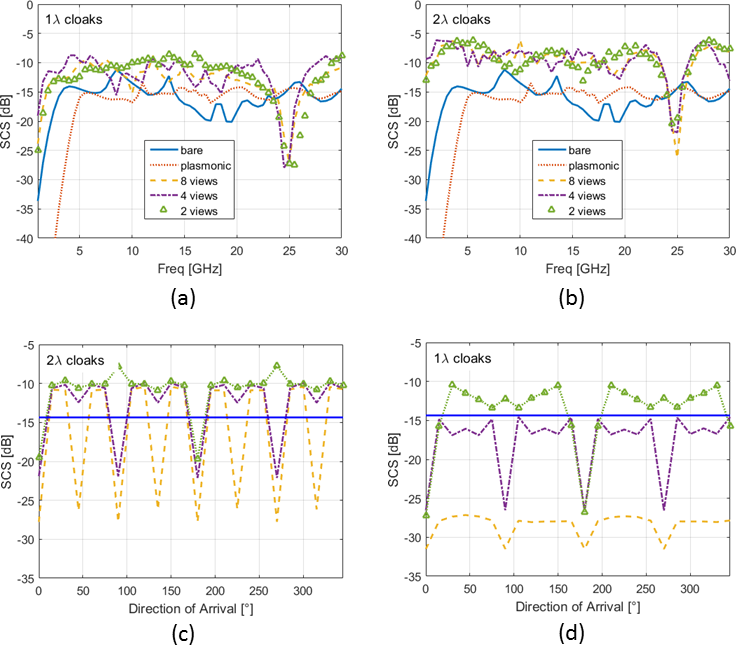}
\caption{\small SCS calculated at fixed DoA as a function of the frequency for (a) 1$\lambda$ and (b) 2$\lambda$ dielectric cloaks and comparison with the bare object and plasmonic cloak: continuous line (bare uncloaked object), dotted line (plasmonic cover), upward-pointing triangle line ($2$-views dielectric cloak), dash-dot line ($4$-views dielectric cloak) and dashed line ($8$-views dielectric cloak);
(c)-(d) SCS calculated at 25 GHz as a function of the DoA, same legend as in (a)-(b).}
\label{fig:cloak_SCS}
\end{figure}

As shown in Fig. \ref{fig:cloak_SCS}(a), in subwavelength condition, the plasmonic cover (dotted line) drastically reduces the scattering levels with respect to the uncloaked case (continuous line) and, at the same time, its cloaking response at $4$ GHz is of omnidirectional kind, as shown in Fig. \ref{fig:cloak_SCS}(b), i.e., completely flat with respect to change in the DoA. 

For what concerns the ordinary dielectric cloaks with radius $b=1.2$ cm, in Fig. \ref{fig:cloak_SCS}(a), they all share the minimum around the designed frequency $25$ GHz, but they possess very different angular responses as reported in Fig.  \ref{fig:cloak_SCS}(c), due to the fact that they have been designed for different incoming waves. In the case $b=1\lambda$, the cloaking effect is clearly affected by changing the DoA: for the $2$-views ordinary dielectric cloak (upward-pointing triangle line), the performance becomes worse when non-optimal DoA is considered, whereas, as expected, two different minima (as the numbers of views for which it has been designed) appear at DoA$=0,\pi$. The $4-$views ordinary dielectric cloak (dash-dot line) improves its omnidirectional performance, ensuring an overall response which is always below the uncloaked case around $-14$ dB (continuous line), especially when the $4$ minima occur. The $8-$views ordinary dielectric cloak (dashed line) shows the best omnidirectional performance, ensuring an overall response between $-27$ dB and $-32$ dB.

In Fig.\ref{fig:cloak_SCS} (b)-(d), the SCS$(\theta_\nu,\omega)$ is shown for the case $b=2.4$ cm ($2\lambda$ at $25$ GHz).
As reported in Fig. \ref{fig:cloak_SCS}(b), the scattering from the ordinary dielectric cloak is larger in the low frequency window with respect their compact counterpart in Fig. \ref{fig:cloak_SCS}(a), due to the fact that now their size is increased: around $25$ GHz, they show a minimum with slightly worse performance in terms of dB reduction compared to the cloaks with $b=1\lambda$ in Fig. \ref{fig:cloak_SCS}(a). Even the angular swing between optimal and worst DoA value, as reported in Fig. \ref{fig:cloak_SCS}(d), is increased. In this $b=2\lambda$ case, the cloaking effect is still affected by changing the DoA, but in a different way with respect to the previous case. The $2$-views ordinary dielectric cloak (upward-pointing triangle line) has wide regions between the minima points DoA$={0,\pi}$ for which it scatters more than the bare object. In this case, also the behavior of the  $4-$views ordinary dielectric cloak (dash-dot line) becomes worse, with several values of the SCS at $25$ GHz above the uncloaked case (continuous line) of about $+5$ dB. Even the $8-$views ordinary dielectric cloak (dashed line) loses its omnidirectionality for a degradation in its performances, whereas ensuring all its $8$ minima around the $-27$ dB value.  


%

\section*{Conclusion}\label{sec:6}

The synthesis of all dielectric cloaks has been tackled through the solution of an inverse scattering problem where zero scattered field is properly pursued with artificial and natural dielectric materials within and outside the quasi-static frequency regime, respectively.
It has been found and discussed that it is not strictly necessary the use of ENZ or ENG materials for cloaking beyond quasi-static limit, completely changing the paradigm suggested by quasi-static formulas and exploring the potentiality of cloaking techniques via scattering cancellation. For this reason, natural dielectric cloaks can be  synthesized in a relatively compact and easy fashion. Moreover, the behavior of the cloak is non-resonant and a non-negligible operational bandwidth (at $3$ dB) can be achieved (about $15\%$). 
Interestingly, being made of only natural dielectrics substances, the cloak could be suitably manufactured by means of solid multi-filament printing techniques when the permittivity values can be possibly achieved using alternative dielectric mixtures with differing volume fractions and particle sizes \cite{LaSpada2016,Valentine2009}.
As a result, this would pave the way to cloaks easy to fabricate with respect to natural dielectrics available in the frequency range of interest and exploiting non-uniqueness of solution in the synthesis procedure based on inverse scattering techniques. 
We are currently working on the realization of such non-homogeneous dielectric covers for experimental measurement of their performance. Further efforts will be addressed to take into account additional design parameters and constraints, as well as to tackle the full 3D electromagnetic cases. 
%

%


\section*{Acknowledgements}
The authors would like to thank  Roberta Palmeri for her valuable support in performing the numerical analysis with COMSOL multiphysics.



\end{document}